\RequirePackage{fix-cm}
\documentclass[twocolumn,epjc3]{svjour3}  
\smartqed  
\RequirePackage{mathptmx}      
\RequirePackage[colorlinks,citecolor=blue,urlcolor=blue,linkcolor=blue]{hyperref}
\RequirePackage{amsmath}
\RequirePackage{amsfonts}
\RequirePackage{amssymb}
\RequirePackage{physics}
\RequirePackage{color, soul}
\RequirePackage{subcaption}
\RequirePackage{graphicx}
\RequirePackage{caption}
\RequirePackage{cite}
\RequirePackage{subcaption}

\newcommand{\bx}{{\boldsymbol{x}}}

\journalname{Eur. Phys. J. E}
\begin{document}

\title{Hierarchical microphase separation in non-conserved active mixtures} 
\author{Yuting I. Li, Michael E. Cates} 

\author{Yuting I. Li \thanksref{e1,addr1}
        \and
        Michael E. Cates \thanksref{addr1}
}

\thankstext{e1}{e-mail: yuting.li@damtp.cam.ac.uk}

\institute{ DAMTP, Centre for Mathematical Sciences, University of Cambridge, Wilberforce Rd, Cambridge CB3 0WA, United Kingdom \label{addr1}
     }

\date{Received: date / Accepted: date}

\abstractdc{Non-equilibrium phase separating systems with reactions can break time-reversal symmetry (TRS) in two distinct ways. Firstly, the conservative and non-conservative sectors of the dynamics can be governed by incompatible free energies; when both sectors are present, this is the leading-order TRS violation, captured in its simplest form by `Model AB'. Second, the diffusive dynamics can break TRS in its own right. This happens only at higher order in the gradient expansion (but is the leading behaviour without reactions present) and is captured by `Active Model B+' (AMB+). Each of the two mechanisms can lead to microphase separation, by quite different routes. Here we introduce Model AB+, for which both mechanisms are simultaneously present, and show that for slow reaction rates the system can undergo a new type of hierarchical microphase separation, whereby a continuous phase of fluid 1 contains large droplets of fluid 2 within which small droplets of fluid 1 are continuously created and then absorbed into the surrounding fluid-1 phase. In this state of `bubbly microphase separation' the small-scale 1-in-2 droplets arise by the conservative diffusive dynamics with the larger scale 2-in-1 structure governed by the nonconservative reactions. }

\maketitle

\section{Introduction}
\label{intro}
Recently, it has emerged that the non-membrane compartments of cells, also known as biomolecular condensates, can be fruitfully viewed as phase separated liquid-liquid mixtures undergoing chemical reactions \cite{hyman2014liquid, berry2018physical, Jacobs2017biophysical, Weber2019review, hyman2011, banani2017, brangwynne2015}. In contrast to passive binary fluids, cells continuously consume fuel, driving the system away from equilibrium. Put differently, the liquid mixtures within cells are examples of active matter. The characteristics of phase separation in these biomolecular condensates are distinct from equilibrium counterparts, in that microphase separation is often observed without any energetic cause \cite{Weber2019review}. (In equilibrium, the latter requires long-range interactions mediated by charged species or block copolymers, for example. \cite{Muratov2002}) This echoes studies of bacteria colonies whose phase-separation (coarsening) is arrested by birth-and-death dynamics to give patterns on a finite length scale \cite{catesPNAS2010}. This suggests that, like the latter, the physics of biomolecular condensates can be captured schematically by `active field theories' based on minimal, $\phi^4$-type ingredients \cite{Wittkowski_NatComms, Tiribocchi_PRL, TjhungPRX2018, Singh_PRL,modelAB,cates2018JFM}.

To construct top-down active field theories for particular systems, one can start with the desired symmetry and conservation laws, and write down the lowest order theory. This follows lines similar to Hohenberg and Halperin's classification of equilibrium models \cite{HH}. In the present work we focus on theories for a single scalar field $\phi$, which could be (for bacteria or swimming organisms) a rescaled density, or (for the binary mixtures of interest here) a chemical composition variable. Within the resulting class of minimal scalar models, several mechanisms have been found to arrest phase separation. In our previous work, we explored the lamellar phase and the droplet suspension phase in Model AB \cite{modelAB}. The latter is a canonical representation of phase-separating conservative dynamics coupled to non-conservative reactions, governed by Model B and Model A respectively, in such a way that the underlying free energy structures in each sector are incompatible. This breaks detailed balance and hence time reversal symmetry (TRS).  The finite length scale of the resulting microphase separation (between regions rich in species 1 and 2 respectively) arises from the balance between the reactions (which convert the majority species into the minority) within the two separated phases, and the conserved, diffusive currents that transport species across the interfaces from high to low chemical potential. In steady state, the interfaces cannot be too far apart, otherwise the buildup of minority species within each phase would cause a spinodal instability and create new interfaces. Microphase separation, rather than bulk phase separation, occurs whenever the stable fixed point composition of the homogeneous reaction dynamics lies between the binodals of the conservative phase separation, so that the uniform bulk phases are chemically unstable \cite{catesPNAS2010,modelAB,ZwickerPRE}. 

Meanwhile, phase separated states have also been reported in fully conservative active systems described by Active Model B+ (AMB+). This model is constructed by adding a complete set of lowest-order non-equilibrium current terms to the standard (passive) Model B \cite{HH} (with differing outcomes, in more than one dimension, from an earlier, less complete, Active Model B \cite{Wittkowski_NatComms}) \cite{TjhungPRX2018}. The active terms in AMB+ can send into reverse the Ostwald process for droplets, which would normally lead to full phase separation by causing small droplets to shrink and disappear while large ones grow \cite{cates2018JFM}. The reverse Ostwald process, which formally \\
emerges through a negative effective interfacial tension in the expression for the Laplace pressure at a droplet surface, instead equalises the droplet sizes. (Note that other definitions of interfacial tension do not become negative, so that droplets remain stably spherical for example.) In the absence of noise, this mechanism creates a set of uniformly sized droplets whose number is governed by the initial condition. However, when noise is present, this ultimately selects the length scale for microphase separation by a subtle interplay of various effects. (Those effects include noise-induced nucleation of small droplets and coalescence of large ones, coupled to a collective homogenization of droplet size caused by reverse Ostwald \cite{TjhungPRX2018}.) 

As just described, Model AB and AMB+ each involve a single scalar field on its own. More elaborate active field theories can be constructed by coupling a compositional scalar to a momentum-preserving fluid flow, along lines establised in passive systems as `Model H' \cite{HH}. It is known that this coupling can give further distinctive routes to microphase separation, for example when activity in effect reverses the sign of the interfacial tension that governs the stress exerted on the fluid at the boundary of the droplet (but without doing the same for the Ostwald currents). This happens when the droplet is made of so-called `contractile' material \cite{Tiribocchi_PRL, Singh_PRL}. In this paper we omit these further complications, and consider only how to combine the types of physics already captured by Model AB and AMB+, in which there is no separate momentum conservation, and the reaction-diffusion dynamics of $\phi$ give a complete description of the pertinent physics. 

In our previous work on Model AB \cite{modelAB}, we broke TRS only at the lowest possible order in $(\grad, \phi)$. This entails choosing chemical potentials for the Model B and Model A sectors that could each give an equilibrium system in the absence of the other, so that $\mu_\mathrm{A,B} = \delta F_\mathrm{A,B}/\delta\phi$, but also choosing $F_\mathrm{A}\neq F_\mathrm{B}$. This leads to phase separation via the mechanism described previously. 
However, in a given physical system, once TRS is broken in this fashion it is likely to be broken elsewhere as well. In particular, we know that the active currents represented by the $\phi$-conserving AMB+ (which do not derive from any local chemical potential, of the form $\delta F/\delta \phi$ or otherwise), can independently cause phase separation via the reverse Ostwald process \cite{TjhungPRX2018}. Even if these are formally subdominant when expanding in $(\grad, \phi)$, such terms can in principle interact in a strongly non-additive way with the microphase separation in Model AB, particularly if parameters are chosen so that the two mechanisms act on well-separated length- and time-scales.

Accordingly, in this paper we construct Model AB+ which includes both types of TRS breaking at once, and use it to give a first account of the interplay between the two mechanisms of microphase separation. Our aim in this brief and exploratory study is not to comprehensively explore the parameter space, but rather to show proof of principle that new physics can indeed emerge from this combination of mechanisms. In particular, in the case where the chemical reactions are slow, we will give numerical evidence for a hierarchical phase separation in which the conservative and non-conservative mechanisms act simultaneously at small and large length-scales respectively. This results in a new, dynamical steady-state structure that we call `bubbly microphase separation'.

The paper is organised as follows. In Section \ref{modelAB+}, we summarise the original Model AB as presented in our previous work \cite{modelAB} and recall its various stationary patterns before introducing Model AB+. Section \ref{steady_states} then explores some steady states of Model AB+ with comparisons to the corresponding Model AB pattern without the active current terms from AMB+. These include droplet emulsions alongside the bubbly microphase-separated state. We then present in Section \ref{emulsion} a mean-field Ostwald-type calculation for growth of a $\phi>0$ (liquid) droplet in a bath with $\phi<0$ (vapour). This explains a transition (or crossover) between statistically different droplet emulsion states upon increasing the target density of the chemical reactions above the dilute binodal of the conservative phase separation. Section \ref{conclusion} briefly summarises these findings and suggests directions for future work. 

\section{Model AB+}
\label{modelAB+}
We will first briefly introduce both Model AB and Active Model B+, before combining their two dynamics to construct Model AB+. 
Model AB, constructed as a combination of Model B and Model A, is a minimal field theory for a scalar composition variable $\phi(\bx)$, in reaction-diffusion systems in which the conservative part of the dynamics, taken alone, would give phase separation \cite{modelAB}: 
\begin{equation}
\begin{split} 
\partial_t \phi &= - \grad \cdot \boldsymbol{J}- M_\mathrm{A} \mu_\mathrm{A} + \sqrt{2 \epsilon M_\mathrm{A}} \Lambda_\mathrm{A} \\
\boldsymbol{J} &=  - M_\mathrm{B} \grad \mu_\mathrm{B} + \sqrt{2 \epsilon M_\mathrm{B}} \boldsymbol{\Lambda}_\mathrm{B} \\
\mu_\mathrm{B} &=  - \alpha \phi + \beta \phi^3 - \kappa \nabla^2 \phi  \\
\mu_\mathrm{A} &= u (\phi-  \phi_\mathrm{a})  (\phi - \phi_\mathrm{t}) 
\end{split} 
\label{eq:one}
\end{equation}
Here $M_\mathrm{A, B}$ are mobility constants, $\Lambda_\mathrm{A}, \boldsymbol{\Lambda}_\mathrm{B}$ are spatiotemporal white noises and $\alpha, \beta, \kappa, u$ are positive constants. The parameter $\epsilon$ is temperature, or an equivalent noise parameter in cases where the primary noise source is not thermal. We also assume that $\phi_\mathrm{a} < \phi_\mathrm{t}$. Here $\phi_\mathrm{a}$ is the absorbing state composition of the reaction diffusion dynamics (this would be the zero density state for a birth-death system), which marks the physical lower limit for $\phi$, whereas $\phi_\mathrm{t}$ is the `target density' of the chemical reactions, such that there is a stable fixed point of the dynamics at $\phi(\bx) = \phi_\mathrm{t}$, {\em if} the diffusive tendency to phase separate is switched off. 

Given the form of chemical potential $\mu_B$ chosen here, if we had chosen instead $\mu_\mathrm{A} = \mu_\mathrm{B}$, the conservative and nonconservative dynamics would share a free energy $F_\mathrm{B} = \int \left(-\alpha\phi^2/2+\beta\phi^4/4+\kappa(\nabla\phi)^2/2\right)\,\mathrm{d}\bx$ so that the model would be an equilibrium one. 
Much less obviously, but as discussed in our previous paper \cite{modelAB}, if $\mu_\mathrm{A}$ is not equal to $\mu_\mathrm{B}$ but differs from it only by terms linear in $\phi$, the system maps to an equilibrium model with long-range attractions. Hence, by virtue of the quadratic term in $\mu_\mathrm{A}$ (alongside the absence of the cubic term), equation (\ref{eq:one}) admits the lowest order mismatch in $\mu_\mathrm{A}$ and $\mu_\mathrm{B}$ that cannot be incorporated into an equilibrium model by judicious matching of higher order terms. 

If the reactions are switched off ($M_\mathrm{A}\to 0$), the diffusive sector of the dynamics drives the system towards bulk phase separation with the two coexisting phases at the binodal densities, $\phi = \pm \phi_\mathrm{B} = \pm \sqrt{ \alpha /\beta}$, as long as the mean density ($\bar{\phi} = \frac{1}{V} \int \mathrm{d} \bx \, \phi (\bx)$ ) is between the binodals. With reactions, broadly speaking, if the target density $\phi_\mathrm{t}$ of the non-conservative sector is between the binodals, the system exhibits micro-phase separation, where material is created in a dilute region, transported across the interface by the Model B current, and eventually destroyed in a dense region. The morphology depends on the value of $\phi_\mathrm{t}$, as shown in the upper panels of figure (\ref{fig:patterns}). For $\phi_\mathrm{t}$ close to zero, the system shows lamellar patterns; otherwise, we either see a droplet phase or its inverse, a bubble phase, depending on which binodal $\phi_\mathrm{t}$ is closer to. In this paragraph, and below, we refer to the two bulk densities as though they were liquid ($\phi>0$) and vapour ($\phi<0$) although of course for biomolecular condensates $\phi$ is composition variable describing whether a fluid mixture of species 1 and 2 is composed mainly of species 1 ($\phi>0$) or species 2 ($\phi<0$). In this context, a rightward current of `material' $\phi$ means a combination of a rightward current of species 1 and a compensating negative current of species 2 at constant total mass density. 

Model AB breaks time reversal symmetry solely by having mismatched free energies in the nonconserved (Model A) and conserved (Model B) sectors. But of course, each sector can, in principle, break detailed balance on its own. In the Model A sector this requires gradient terms in $\mu_\mathrm{A}$ that do not stem from a free energy. At leading order in the gradient expansion, the resulting `Model A+' will not lead to microphase separation under steady-state conditions because, without a conservation law, there are no interfaces in the steady state at which such terms could be large. Instead we expect relaxation towards a uniform state at zero $\mu_\mathrm{A}$, or $\phi = \phi_\mathrm{t}$. These active `A+' terms can of course modify the microphase separation caused by the competing conserved and non-conserved dynamics in Model AB, but we don't expect them to introduce a second type of microphase separation. 

In what follows, we therefore focus on enhancing Model AB by adding active gradient terms to the conservative sector where such terms can lead to microphase separation in their own right. The terms in question are those of Active Model B+ \cite{TjhungPRX2018},
\begin{equation}
\begin{split} 
\partial_t \phi &= - \grad \cdot \boldsymbol{J} \\
\boldsymbol{J} &= M_\mathrm{B} \left [ - \grad \left (  \mu_\mathrm{B} + \lambda \left | \nabla \phi \right |^2 \right ) + \zeta ( \nabla^2 \phi) \grad \phi  \right ]  \\
& \qquad  + \sqrt{ 2 \epsilon M_\mathrm{B} } \boldsymbol{\Lambda}_\mathrm{B} 
\end{split} 
\end{equation}
where $\mu_\mathrm{B}$ is the same as before. The $\lambda$ and $\zeta$ terms are the lowest order Landau-Ginzburg terms that break TRS in systems with $\phi$ conservation. (One other term at this order can be absorbed into a $\phi$-dependent square-gradient coefficient $\kappa(\phi)$ without breaking the free energy structure.)
It has been shown that the $\lambda$ and $\zeta$ terms not only shift the binodals, but that the latter can also lead to effectively negative surface tensions for either liquid droplets ($\phi>0$) or vapour bubbles ($\phi<0$), albeit not both at once,  depending on the sign of $\zeta$ \cite{TjhungPRX2018}. The tension in question is the one governing Laplace pressures and hence Ostwald ripening, not other physics (such as fluctuations at the interface between phases which remains stable). The resulting reverse Ostwald process causes microphase separation into an emulsion of finite droplets that do not coarsen, which is of course not seen in equilibrium Model B. In renormalization-group terms, this microphase separation appears to be connected with a strong coupling regime at large enough values of $(\lambda, \zeta)$. As such, these variables remain important even though, being higher order in the Landau-Ginzburg expansion, the are formally irrelevant in the neighbourhood of the Wilson-Fisher fixed point which controls the critical onset of bulk phase separation \cite{caballeroRG}.

This suggests that equally important physics could also be lost by ignoring these non-integrable gradient terms in the conservative sector for systems with both conserved and non-conserved dynamics, whereas we did ignore them when constructing Model AB as a canonical model for that case \cite{modelAB}. 
In light of this, we now construct Model AB+ by adding the $\lambda$ and $\zeta$ terms to the conservative sector of Model AB, so that \eqref{eq:one} is replaced by:
\begin{equation}
\begin{split} 
\partial_t \phi &= - \grad \cdot \boldsymbol{J}- M_\mathrm{A} \mu_\mathrm{A} + \sqrt{2 \epsilon M_\mathrm{A}} \Lambda_\mathrm{A} \\
\boldsymbol{J} &=   M_\mathrm{B} \left [ - \grad \left (  \mu_\mathrm{B} + \lambda \left | \nabla \phi \right |^2 \right ) + \zeta ( \nabla^2 \phi) \grad \phi  \right ]   \\
& \qquad + \sqrt{ 2 \epsilon M_\mathrm{B} } \boldsymbol{\Lambda}_\mathrm{B}  \\
\mu_\mathrm{B} &=  - \alpha \phi + \beta \phi^3 - \kappa \nabla^2 \phi  \\
\mu_\mathrm{A} &= u (\phi-  \phi_\mathrm{a})  (\phi - \phi_\mathrm{t}) 
\end{split} 
\label{eq:phi_dot_ab}
\end{equation}

The rest of this paper addresses Model AB+ as written in the form \eqref{eq:phi_dot_ab}. All numerical simulations are performed in d=2 with periodic boundary conditions. Spatial derivatives are computed in real space following the methods detailed in Appendix A of \cite{TjhungPRX2018}. Time integration is performed using the explicit Euler-Maruyama method \cite{kloeden}. Throughout the paper, unless otherwise stated, we set $M_\mathrm{B} = 1, \alpha = \beta = 0.25, \kappa = 1, u = ( - \phi_\mathrm{a} + \phi_\mathrm{t} /2 ), \phi_\mathrm{a} = -10$. Note that we make no attempt at a systematic exploration of parameter space, but focus on selected regimes where new physics can be expected by adding nonzero $\lambda,\zeta$ terms to parameter sets that we used previously to study Model AB. (In particular, the somewhat arbitrary choice for $u$ was adopted in our previous paper \cite{modelAB} and is kept here for consistency. In principle, one can simply set $u=1$ and the results will be qualitatively the same.) 

\section{Steady states} 
\label{steady_states}
Model AB+ has at most one stationary state of uniform density, which lies at the target density $\phi_\mathrm{t}$. (Any other uniform state has $\dot\phi\neq 0$ from the reactions, with no diffusive currents that could balance this.) Linear stability analysis finds this uniform state to be unstable to spatial perturbations if 
\begin{equation}
M_\mathrm{B} \frac{\tilde{\alpha}^2}{2 \kappa } > M_\mathrm{A} \tilde{u} 
\end{equation}
where $\tilde{\alpha} =  2 \beta \phi_\mathrm{t}^2 - \alpha$, $\tilde{u} = u(-\phi_\mathrm{a} + \phi_\mathrm{t})$. The two sides of the inequality represent characteristic relaxation rates via diffusion and reactions, respectively, on length scales set by $\kappa$. In many biological situations, relatively rapid thermal diffusion of chemical species is accompanied by a relatively slow reaction-driven turnover time, so that this `slow reaction' condition is easily obeyed. Hence in this paper we focus on the unstable regime far away from the threshold of linear instability. We further assume that $\phi_\mathrm{a} \leq - \phi_\mathrm{B}$ so the reaction dynamics is approximately linear in $\phi(\bf x)$ for values lying between the binodals. This means that we have an approximate symmetry $(\lambda, \zeta, \phi) \rightarrow (-\lambda, -\zeta, -\phi)$ in our system, similar to Active Model B+ where this symmetry is exact \cite{TjhungPRX2018}. 

\begin{figure*}
\includegraphics[width=0.8\textwidth]{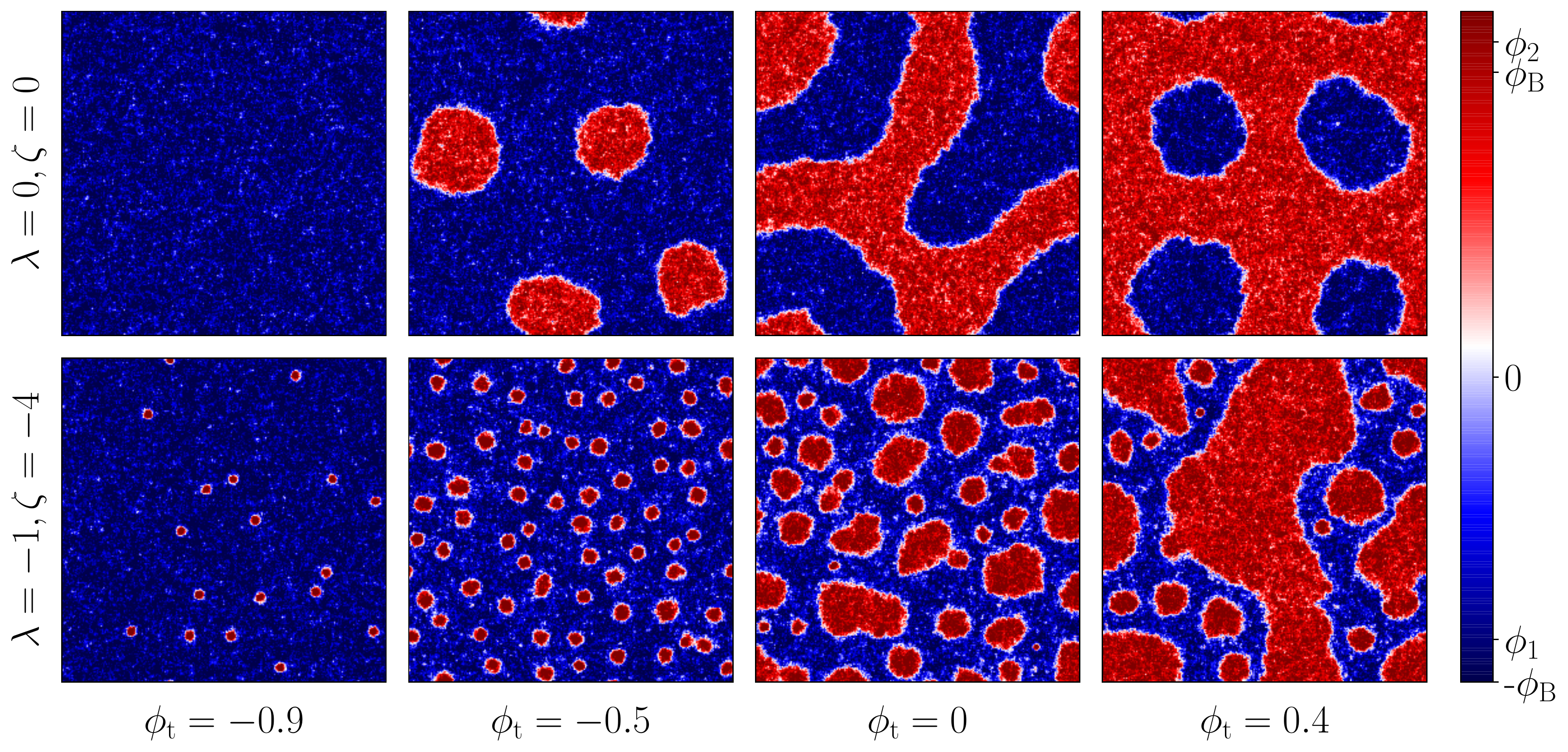}
\caption{Steady state patterns for $\phi_\mathrm{t} = -0.9, -0.5, 0, 0.4$ for two sets of $(\lambda, \zeta)$ values. For $\lambda = 0, \zeta=0$ (as in Model AB), the patterns are stationary modulo noise fluctuations and are shown in the top panel. For $\lambda = -1, \zeta = -4$, the steady states are dynamical with droplets of the dense phase (red) constantly spawn in the dilute phase (blue), snapshots of which are shown in the bottom panel. }
\label{fig:patterns} 
\end{figure*}

Some typical steady states are shown in Fig.~(\ref{fig:patterns}) for $\lambda = -1, \zeta = -4$, where the Ostwald tension for droplets of the dense phase ($\phi>0$) are negative. These are directly compared with the pure Model AB case ($\lambda = \zeta = 0$). For low values of $\phi_\mathrm{t}$, where pure AB is in a uniform state we see an emulsion of these dense (`liquid') droplets in the dilute phase (`vapour', $\phi<0$), stabilized by reverse Ostwald. At high $\phi_t$, chosen so that the Model AB mechanism leads to a microphase-separated state with liquid phase in the majority (hence an emulsion of vapour bubbles in liquid), the active current terms cause phase inversion of this state so that the system is liquid-in-vapour. This is inevitable for sufficiently negative $\lambda,\zeta$ where the conservative dynamics causes microphase separation on a relatively short length scale since, importantly, these terms {\em only} stabilize liquid in vapour droplets and not vice versa (unless their signs are reversed, in which case so is the whole phase diagram). 

Because even at these parameter values the binodals $\phi_{2,1}$ are only modestly shifted from the equilibrium values $\pm\phi_\mathrm{B}$, the global phase volumes of the liquid and vapour phases remain primarily under the control of the reaction dynamics and are hence set by $\phi_t$. At large enough values of this quantity the majority liquid phase unavoidably percolates, but the minority phase of disconnected large vapour bubbles contain within them small liquid droplets that are stabilized by reverse Ostwald. (The interior of such a bubble thus resembles a piece of the liquid-in-vapour emulsion seen at smaller $\phi_t$.) These small liquid droplets are continuously produced within the vapour bubbles but then grow, diffuse and merge into the surrounding liquid phase. This behaviour closely resembles the `bubbly phase separation' reported previously for AMB+, except that in the latter case, there is only a single vapour domain in the system which can effectively be viewed as a bulk phase separation between the liquid-in-vapour emulsion and excess liquid \cite{TjhungPRX2018}. (Note also that \cite{TjhungPRX2018} mainly addresses the case with $\lambda,\zeta>0$ for which the identities of the `liquid' and `vapour' phases are, trivially, interchanged from those in the present discussion.) In the presence of the Model AB mechanism, this type of emulsion/liquid bulk phase separation is itself unstable since the bulk liquid phase is not at the target density. Accordingly, for slow reaction dynamics the system must homogenise once again at some larger scale so that the chemical conversion of the majority into the minority species in each neighbourhood can be balanced by diffusive mass transport. Accordingly we see a finite density of the large liquid-in-vapour emulsion bubbles whose size cannot grow further. Echoing the language of \cite{TjhungPRX2018}, we refer to this state as `bubbly microphase separation'. 

\begin{figure*} 
\includegraphics[width=0.8\textwidth]{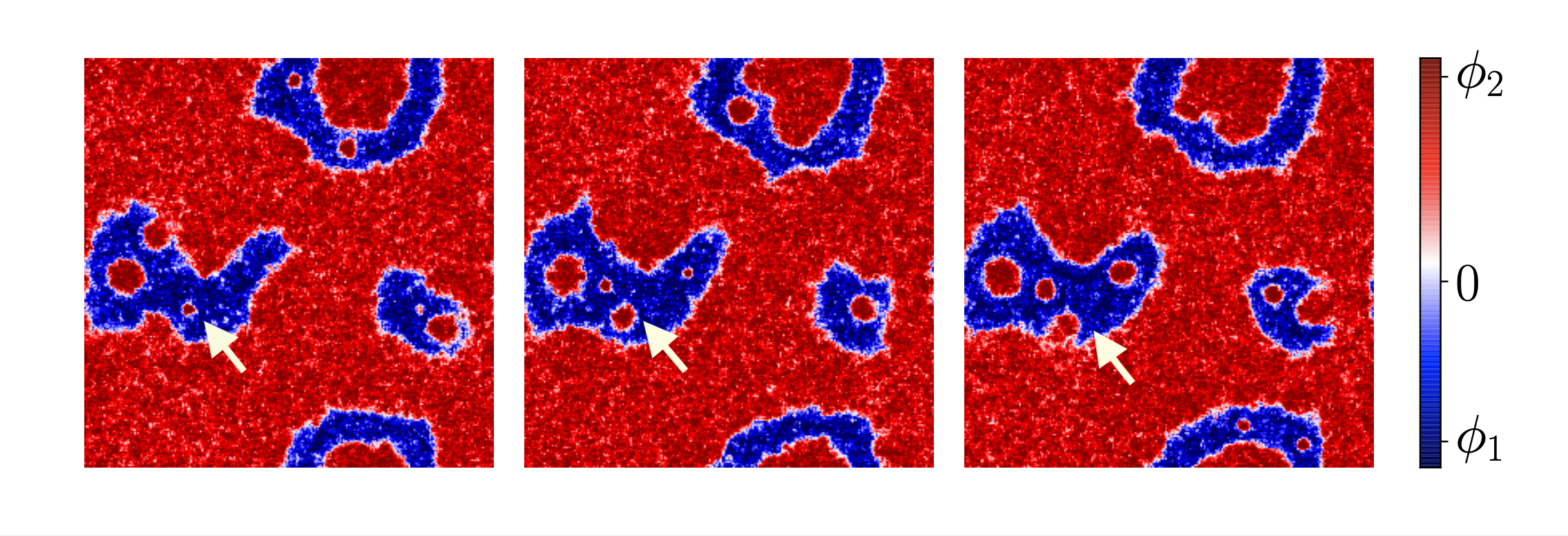}
\caption{Three snapshots of the life-cycle of a dense droplet created inside the dilute phase, as indicated by the arrow. The simulation parameters are $uM_\mathrm{A} = 1 0^{-6}, \phi_\mathrm{t} = 0.6$.}
\label{fig:snapshots}
\end{figure*}

A snapshot of the life-cycle of droplets is shown in figure (\ref{fig:snapshots}) showing the growth of a small nucleated droplet, its growth, and coalescence into the surrounding liquid phase. As stated previously this echoes the findings of \cite{TjhungPRX2018} for the life cycle of liquid droplets within a bulk phase separation between an emulsion of such droplets and excess liquid.

\section{Ostwald Dynamics in Model AB+} 
\label{emulsion} 
In the mean field limit, the binodals of AMB+, defined as the steady state coexisting densities for a flat interface, can be computed analytically \cite{Tiribocchi_PRL}. We will not show the calculation here but merely quote the results for $\lambda = -1, \zeta = -4$: the dense binodal shifts to $\phi_1 = 1.10$ and the dilute binodal shifts to $\phi_2 = - 0.86$. These replace the values $\pm\phi_\mathrm{B} = \pm 1$ for the passive model (with $\alpha = -\beta$ as chosen above). 

In addition, around circular two-dimensional droplets, the density $\phi$ close to the interface is further modified by an offset depending on the radius of the droplet $R$. In equilibrium Model B, this curvature induced shift around a liquid droplet is proportional to $\sigma/R$, where $\sigma$ is the surface tension. As shown in \cite{TjhungPRX2018} for AMB+ the surface tension defined this way can be negative for sufficiently negative of $\lambda$ and $\zeta$, thus reversing the Ostwald ripening for droplets, arresting their growth at a finite size. If we now switch on the chemical reactions to give Model AB+, then so long as these reactions are slow (which is the regime addressed in this paper) the dynamics on the length-scale of the interface width $\xi_0 \simeq \sqrt{\kappa/2\alpha}$ is dominated by the conservative current. This means that we can carry across from AMB+ the results for the curvature-induced offset to the binodals. 

Consider N liquid droplets of radius $R_j \gg\xi_0$ for $j \in (1, 2, \dots, N)$ in a dilute bath in a domain of volume $V$ with periodic boundary conditions. We now single out one droplet of radius $R_i$ and treat the rest as homogeneous background. Let $\phi_\pm(\boldsymbol{x})$ be the composition field inside and outside the singled-out droplet respectively, and define variables $\psi_\pm(\boldsymbol{x})$ as the deviation from the modified binodal densities: 
\begin{equation}
\begin{split} 
\phi_-(\bx )  = \phi_2 + \psi_- (\bx) \qquad  &   | \bx | \geq R_i \\
\phi_+(\bx)  = \phi_1+ \psi_+ (\bx ) \qquad & 0 \leq | \bx | < R_i
\end{split} 
\end{equation}
For reasons given in \cite{modelAB}, provided that $\phi_\mathrm{t}$ is close to the modified dilute binodal $\phi_2$ (so that there is a large excess of vapour around the droplet), and so long as the timescale separation between conservative and nonconservative dynamics holds (slow reaction regime), we can linearise the deterministic part of equation (\ref{eq:phi_dot_ab}) to obtain, 
\begin{equation}
\begin{split} 
&\partial_t \psi_-  = D_- \nabla^2 \psi_- + (1-\lambda) g^0_- + g^1_- \psi_- + V^{-1} \sum_{j=1}^N 2 \pi R_j  J_-^j  \\
&\partial_t \psi_+  = D_+ \nabla^2 \psi_+ + g^0_+ + g^1_+ \psi_+
\end{split} 
\label{eq:linear}
\end{equation}
where $D_\pm = M_\mathrm{B} (  - \alpha +3 \beta (\phi_{1, 2} )^2) $ and we have defined $g(\phi) = - M_\mathrm{A} \mu_\mathrm{A} (\phi), g^0_\pm = g( \phi_{1, 2} ) , g^1_\pm = g'(\phi_{1, 2} ), \lambda=V^{-1} \sum_j \pi R_j^2$ for conciseness. In the last term in the equation for $\psi_-$, $J_-^j$ denotes the current at the surface of the $j$th droplet and thus this term accounts for the injection of mass by all the droplets. Note that we neglect the $\kappa$ term here as $\psi$ varies on a length-scale much larger than the interfacial width. The linearised equations need to be solved with the appropriate boundary conditions: at $|\bx| \rightarrow 0$ and $|\bx| \rightarrow \infty$, we only need $\psi_\pm$ to be finite; at the boundary of the droplet, we require $\psi_\pm = \delta_\pm(R)$, where $\delta_\pm (R)$ is the aforementioned offset for the dense and dilute phase respectively. 

We will now sketch the perturbative Ostwald calculation with results presented only at key steps and refer to Sec.\ 3.2 of \cite{modelAB} for more detailed explanations of the scheme. This is inspired by the standard Ostwald calculation for both passive and active systems \cite{TjhungPRX2018, cates2018JFM} and its adaptation to multiple droplets \cite{modelAB, ZwickerPRE}, but we implemented it here, for the first time, to include {\em both} the nonconservative reaction terms in the bulk phases (following \cite{modelAB}) {\em and} the effects of active currents on the matching conditions (following \cite{TjhungPRX2018}) which cause reversal of the Ostwald dynamics. 

\begin{figure*} 
\centering
\begin{subfigure}[b]{0.62\textwidth} 
 \includegraphics[width=\textwidth]{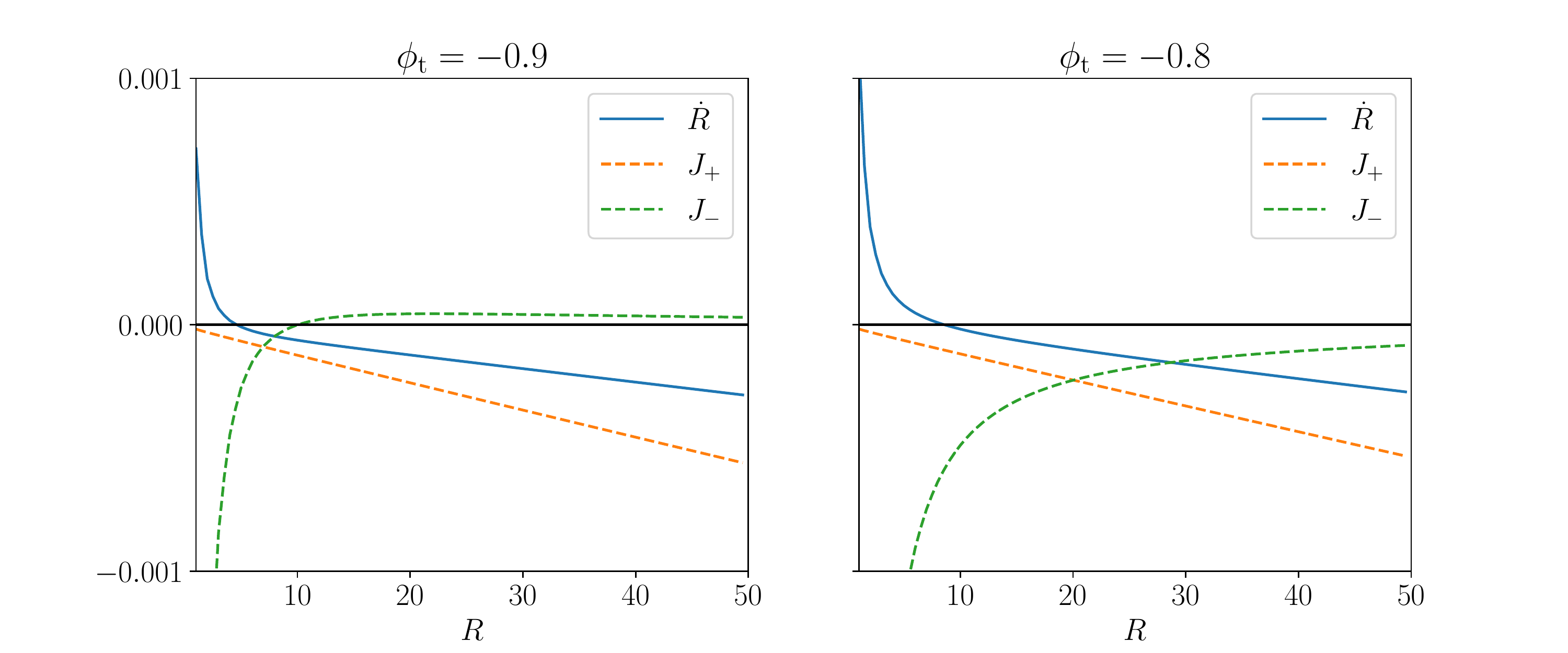}
\caption{}
\label{fig:r_dot}
\end{subfigure} 
\begin{subfigure}[b]{0.35\textwidth}
\includegraphics[width=\textwidth]{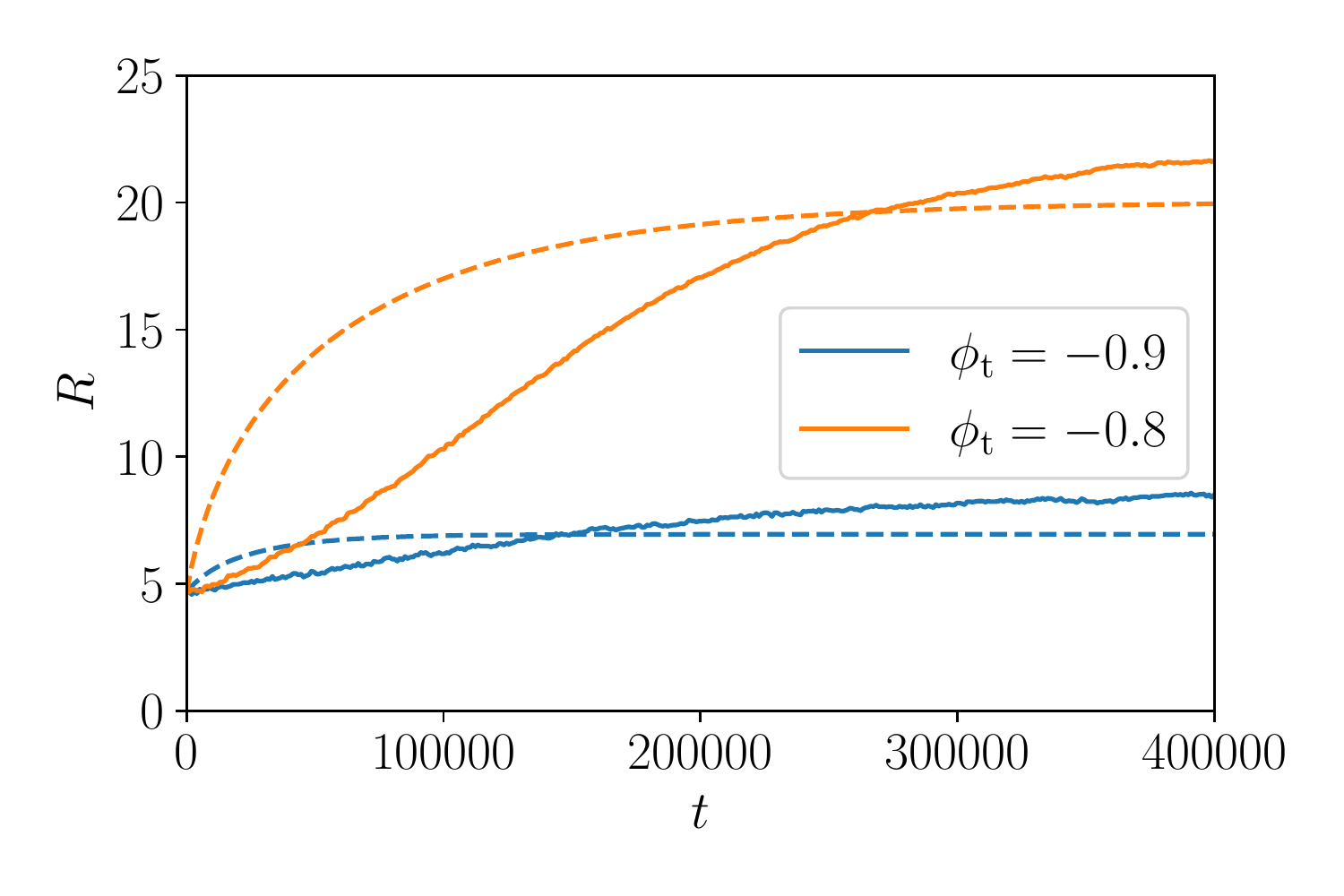}
\caption{} 
\label{fig:comparison}
\end{subfigure} 
\caption{(a) Plots of the growth rate $\dot{R}$ and the currents inside $J_+$ and outside $J_-$ for a dense (liquid) droplet of radius $R$ in a dilute (vapour) bath for $\phi_\mathrm{t} < \phi_1$ and $\phi_\mathrm{t} > \phi_1$ (recall $\phi_1 = -0.86$). In both cases, the $\dot{R}$ curve crosses the x-axis from above, indicating a stable fixed point for $R$. The key difference is that in the left panel the $J_-$ curve is positive as $R \rightarrow \infty$ whereas $J_-$ remains negative on the right panel. (b) Comparisons between theoretical predictions (dotted line) against simulations (solid line) for  $\phi_\mathrm{t} < \phi_1$ and $\phi_\mathrm{t} > \phi_1$. All plots are produced with the same reaction rate $uM_\mathrm{A}= 1\times10^{-6}$. The noise level used in the simulations is $\epsilon = 0.001$.}
\end{figure*}

For fixed droplet radii $\{ R_j \}$, we solve for the stationary state of the linearised equations with $\psi_\pm(R_i) = \delta_\pm (R_i) $, then compute the current $J_\pm^i$ at the interface with the quasi-static solutions as a function of the droplet radii $\{R_j\}$, 
\begin{equation}
\begin{split} 
J_+^i &= D_+ k_+ ( c_+ - \delta_+ ( R_i )  )\frac{ I_1 ( k_+ R_i) }{ I_0 (k_+ R_i ) }\\
J_-^i &= - D_- k_- (c^i_- - \delta_-(R_i) ) \frac{K_1(k_- R_i ) }{K_0 ( k_- R_i ) } \\
c^i_- &= c_- ( 1 - \lambda) - (g_-^1 V)^{-1} \sum_{j=1}^N 2 \pi R_j J_-^j 
\end{split} 
\end{equation}
where $K_{0, 1}, I_{0, 1}$ are the modified Bessel functions, \\
 $k_\pm = \sqrt{-g_\pm^1/D_\pm}$ and $c_\pm =  - g_\pm^0 / g_\pm^1 \approx \phi_\mathrm{t}  - \phi_{2, 1}$. At the surface of the singled-out $i$th droplet (treating the rest as a homogeneous bath), the mismatch of the diffusive currents on the two sides of the interface $J^i_\pm$ shifts the droplet boundary, thus changing the radius of the droplet, 
\begin{equation}
\partial_t R_i =  \frac{ J_+^i( R_i )  - J_-^i  (\{ R_j \} )}{\phi_1 - \phi_2} 
\label{eq:r_dot} 
\end{equation}  
So far, we have reduced the multiple-droplet dynamics to $N$ coupled equations for $\{R_j \}$ by treating each droplet in isolation. One can then immediately spot a solution where all the droplets have the same radius. Let this radius be R, the currents $J_-^i$ can then be obtained explicitly by substituting the $c_-^i$ equation into the $J_-^i$ equation:
\begin{equation}
J_-^i (R) = - D_- k_- \frac{K_1}{K_0} ( c_- ( 1 - \lambda) - \delta_-(R) ) \left ( 1 - \frac{ 2 \lambda D_- k_-}{ g_-^1 R} \frac{K_1}{K_0}  \right ) 
\label{eq:J_minus}
\end{equation}
where the arguments of the modified Bessel functions $K_{1, 0}$ have been omitted to ease notation. Substituting eq. (\ref{eq:J_minus}) into eq.\ (\ref{eq:r_dot}) yields the growth rate $\dot{R}$ as a function of $R$. The resulting $\dot{R}(R)$ curve is plotted in figure (\ref{fig:r_dot}) for one droplet in a box of size $256^2$ for $\phi_\mathrm{t} < \phi_1$ and $\phi_\mathrm{t} > \phi_1$. Both plots show a stable fixed point: smaller droplets grow towards the stable radius, and larger droplets shrink towards it. These facts respectively explain the spontaneous emergence of liquid droplets within the vapour phase (given the presence of noise at short scales) and also the arrested growth of droplets shown in figure (\ref{fig:patterns}). There are, however, some important differences between the two plots in Fig.~(\ref{fig:r_dot}). Observe that the curve for the currents inside the droplet $J_+(R)$ are similar in the two panels, whereas the current in the vapour phase $J_-(R)$ are qualitatively different. In the case where $\phi_\mathrm{t} < \phi_1$, $\psi_-(\bx )  \rightarrow c_- \approx (\phi_\mathrm{t} - \phi_1) < 0 $ as $ |\bx| \rightarrow \infty$  whereas $\psi_-(R) \rightarrow 0$ for large $R$, leading to diffusive currents away from the droplet on the outside. In contrast, for $\phi_\mathrm{t} > \phi_1$, we have positive $\psi_-$ at infinity, thus the current outside the droplet is always towards the droplet. This has profound consequences: for $\phi_\mathrm{t} < \phi_1$, the zero of the $\dot{R}$ has an upper bound at $\delta_-(R) = c_-$, which is independent of the reaction rate $M_\mathrm{A}$; but for $\phi_\mathrm{t} > \phi_1$, there is no upper bound for the stationary fixed-point radius (at least at the mean field level). Comparisons of the above analysis against numerical simulations with low noise are shown in Fig.\ \ref{fig:comparison}: although the perturbative Ostwald calculation does not quantitatively capture the exact time evolution of $R$, due to the quasi-static assumption, the stable radii between theory and simulations agree reasonably well.  

As one increases the number of droplets in the Ostwald calculation (by changing $\lambda = N \pi R^2/V$ in equation (\ref{eq:J_minus})), the stable fixed-point radius decreases, also observed in \cite{modelAB}; though qualitatively, the fixed-point radius is still substantially larger for $\phi_\mathrm{t}>\phi_1$ than $\phi_\mathrm{t}<\phi_1$. This explains why, at fixed noise level, the emulsion of small and rather monodisperse liquid droplets seen in the bottom left panel of Fig.~\ref{fig:patterns} takes on a quite different character in the two bottom centre panels, with relatively large and quite polydisperse droplets. Clearly in this high phase-volume emulsion state the droplets are neither uniform nor well separated, so that any further attempt to rationalize their statistics requires a much better understanding of the role of noise in the steady state kinetics. This lies beyond our scope; as noted previously, the issue is a surprisingly complicated one even in AMB+ where chemical reactions are absent~\cite{TjhungPRX2018}. 

For even larger values of $\phi_\mathrm{t}$, the Ostwald calculation above continues to qualitatively capture the growth of initially small liquid droplets within vapour regions, but as the droplets grow larger, they come into contact and either form or merge into the continuous phase of liquid which is present in the resulting regime of bubbly or hierarchical microphase separation. 

\section{Conclusion} 
\label{conclusion} 
In this paper we have furthered our investigation into scalar field theories for non-equilibrium phase separating systems. Specifically, we have presented an initial study of the effect of having diffusive dynamics that breaks detailed balance on its own, on top of a mismatch of chemical potentials between the conservative dynamics and the non-conservative dynamics. This mismatch is separately a hallmark of broken time reversal symmetry (broken detailed balance) in active phase separation whenever both types of dynamics are present at once. The resulting new model, Model AB+, incorporates the $\lambda$ and $\zeta$ terms, describing active currents in Active Model B+, into Model AB from our previous paper \cite{modelAB} which focuses on the chemical potential mismatch in isolation. The latter causes microphase separation whenever the chemical reactions drive the system towards a target density that lies between the binodals of a conserved, diffusive phase separation. 

This study was motivated by the ability of the active current terms to reverse the classical Ostwald process \cite{TjhungPRX2018}, which is also associated with a distinct strong coupling regime renormalization group calculations \cite{caballeroRG}.  As a result of this, Model AB+ has two distinct mechanistic channels for microphase separation. Our work has shown that these can interact in nontrivial ways, giving two distinct types of emulsion state and also a `bubbly' microphase separation in which the two processes are operative hierarchically on different scales. 
To be more specific, we focused our studies on parameter regimes where the surface tension for droplets of the dense `liquid' phase is negative. For $\phi_\mathrm{t}$ close to the dilute `vapour' binodal $\phi_1$, suspensions of stable droplets of finite size are observed. Ostwald calculations of a single droplet reveal that if $\phi_\mathrm{t} < \phi_1$ there is an upper limit on the droplet size as $uM_\mathrm{A} \rightarrow 0$ determined entirely by the diffusive dynamics; whereas for $\phi_\mathrm{t} > \phi_1$ such an upper limit does not exist. The dynamics become more complicated for multiple droplets, though the qualitative conclusion remains that there are two structurally distinct liquid-in-vapour emulsions, as reflected in the phase diagram. On the other hand, for larger values of $\phi_\mathrm{t}$, the system settles into the bubbly microphase separated state, characterized by large but finite domains with densities at the two binodals, but with droplets of the liquid phase continuously created within the vapour domains before merging with the surrounding fluid. 

Thus far, we explored some interesting behaviour of Model AB+, but made no attempt to systematically scan its multidimensional parameter space. There is, accordingly, plenty of room to find new physics beyond that presented here. For example, it would be interesting to see what is the interplay between the reverse Ostwald dynamics and the limit cycles observed in Model AB \cite{modelAB}. In these cycles, there is no stationary phase separation but a cycle where phase separation leads to a slow reduction in global density until the system rehomogenizes, whereupon the global density reverses until the system phase separates again. (This is possible when the local reaction rate is sufficiently nonlinear in density.)  It is not clear how this oscillation might interact with a conserved dynamics that favours microphase separation, especially when this is itself a highly dynamical process governed by a nontrivial life-cycle for droplets. Finally, as reported separately for Model AB and AMB+, the noise strength $\epsilon$ plays an important role in selecting the number and size of droplets in the emulsion state \cite{modelAB, TjhungPRX2018} and this aspect deserved further study. 

Finally, we have not made any attempt to connect the parameters or mechanisms of Model AB+ directly with any specific microscopic examples of active phase separation, whether within living cells \cite{hyman2014liquid, berry2018physical, Jacobs2017biophysical, Weber2019review}, bacterial colonies \cite{catesPNAS2010}, or elsewhere. This is generally not easy because the model is constructed top-down to have a minimal structure. While the results are generic, only a comparison with more microscopic, bottom-up treatments can determine whether individual parameters are large or small in any particular case. For example, in purely conservative phase separation with ABPs (Active Brownian Particles), it was found that hard-core repulsion leads directly to the $\lambda$ term in AMB+ but that to recover the $\zeta$ term, additional (e.g., soft-core) interactions are also required \cite{TjhungPRX2018}. Despite these difficulties, we believe that the exploration of a generic model such as Model AB+ can suggest mechanistic explanations for structure formation in binary active systems that might otherwise be quite puzzling. A possible example is when there is emergent structure on more than one length scale, as arises in bubbly microphase separation (bottom right panel of Figure~\ref{fig:patterns}).

\begin{acknowledgements}
We thank Cesare Nardini, Rajesh Singh and Elsen Tjhung for valuable
discussions and code donations. YIL thanks the Cambridge Trust and the Jardine Foundation for a PhD studentship. This work was funded in part by by the European Research Council under the Horizon 2020 Programme, ERC grant agreement number 740269. MEC is funded by the Royal Society.
\end{acknowledgements}

\bibliographystyle{spphys}       
\bibliography{EPJ}   

\end{document}